\documentclass[11pt,a4paper]{article}
\usepackage[T1]{fontenc}
\usepackage{lmodern}
\usepackage{amsmath,amssymb,amsthm,mathtools,mathrsfs}
\usepackage[margin=27mm,top=24mm,bottom=24mm,headheight=15pt]{geometry}
\usepackage{fancyhdr,xcolor,hyperref}
\definecolor{linkink}{RGB}{28,63,85}
\hypersetup{colorlinks=true,linkcolor=linkink,citecolor=linkink,urlcolor=linkink,
  bookmarksnumbered=true,pdftitle={The Core and the Nucleolus of S-T Transversal Games},
  pdfauthor={Pengfei Liu, Han Xiao, Qizhi Fang},
  pdfsubject={Cooperative games and combinatorial optimization},pdflang={en}}
\theoremstyle{plain}
\newtheorem{theorem}{Theorem}[section]
\newtheorem{lemma}[theorem]{Lemma}
\newtheorem{proposition}[theorem]{Proposition}

\theoremstyle{definition}
\newtheorem{example}[theorem]{Example}

\theoremstyle{remark}

\numberwithin{equation}{section}
\renewcommand{\theequation}{\ifnum\value{section}=0 \arabic{equation}\else\thesection.\arabic{equation}\fi}
\newcommand{\one}{\mathbf{1}}
\newcommand{\Bcal}{\mathcal B}
\newcommand{\Core}{\mathcal C}
\DeclareMathOperator{\conv}{conv}
\DeclareMathOperator{\Fix}{fix}
\title{The Core and the Nucleolus\\[2pt]of $S$-$T$ Transversal Games}
\author{Pengfei Liu\qquad Han Xiao\qquad Qizhi Fang\\[4pt]
\small School of Mathematical Sciences, Ocean University of China}
\date{}
\makeatletter
\renewcommand{\@maketitle}{%
  \begin{center}
    {\LARGE\bfseries \@title\par}
    \vspace{10pt}
    {\large \begin{tabular}{c}\@author\end{tabular}\par}
  \end{center}}
\makeatother
\begin{document}
\maketitle

\begin{abstract}
We introduce the $S$-$T$ transversal game on a graph with vertex classes $S$ and $T$. The players are the edges, and the value of a coalition is the maximum number of pairwise disjoint $S$-$T$ connectors contained in this coalition. We characterize the nonempty core as the convex hull of the normalized cut vectors of tight subpartitions. We give polynomial time algorithms for testing core nonemptiness, finding a core allocation, and testing core membership. When the core is nonempty, we give a polynomial time combinatorial algorithm for computing the nucleolus.

\end{abstract}


\section{Introduction}

Let $S$ and $T$ be the two classes of a partition of the vertex set of an undirected graph. An $S$-$T$ connector is a set of edges whose spanning subgraph has a vertex of each class in every component. The connector packing problem asks for the maximum number of pairwise disjoint connectors~\cite{KeijAlgorithm,KS}. Each connector is a transversal of the cuts determined by nonempty vertex sets contained in $S$ or in $T$. We call the associated cooperative game the \emph{$S$-$T$ transversal game}. Each player owns an edge, and the value of a coalition is the maximum number of disjoint connectors its members can form. When one class consists of a single vertex, the minimal connectors are the spanning trees, and the game reduces to the network strength game~\cite{BB}. When every edge joins $S$ to $T$, the connectors are precisely the edge covers of the bipartite graph~\cite{KS}.

For example, suppose that $S$ is a set of suppliers and $T$ a set of customers. A contract requires each supplier to have a route to some customer and each customer to have a route to some supplier. If each contract pays one unit and each link can serve at most one contract, the value of a coalition is the revenue its link owners can earn together.

A core allocation divides the revenue of the grand coalition so that each coalition receives at least its own value. The core is nonempty precisely when the game is balanced~\cite{Bond,ShapBal}. The core and its relaxations have been studied for assignment games~\cite{AssignCore}, linear production games~\cite{Owen}, flow games~\cite{KZ}, and other combinatorial optimization games~\cite{DIN}.

When the core is nonempty, the nucleolus~\cite{Schmeidler} selects a unique allocation in the core. Maschler's scheme computes the nucleolus through a sequence of linear programs that can have exponentially many coalition constraints~\cite{MPS}. The computation of the nucleolus has been studied for tree games~\cite{Tree}, matching games~\cite{UM,WMatch}, fractional matching games~\cite{Cover}, flow games~\cite{FlowHard}, shortest path games~\cite{ShortPath}, network strength games~\cite{BB,EE}, arboricity games~\cite{EE,Arbor}, and matroid packing games~\cite{MP}.

We first show that the core is nonempty if and only if the maximum connector packing value equals the minimum fractional transversal value of the connector family. When these values agree, the core is the convex hull of normalized cut vectors obtained from tight subpartitions of $S$ or of $T$. We then use exchanges of a single connector in a maximum packing to construct two partial orders on blocks of edges. A core allocation gives each edge the sum of the prices of its two blocks. When the core contains more than one allocation, the two orders determine the unique optimal allocation of the second linear program in Maschler's scheme and hence the nucleolus.

\section{Preliminaries}

\subsection{Cooperative games}
A cooperative profit game $(N,v)$ consists of a finite player set $N$ and a characteristic function $v:2^N\to\mathbb R$ with $v(\varnothing)=0$. A subset $C$ of $N$ is a coalition and $v(C)$ is its value. For $x\in\mathbb R^N$, write $x(C)=\sum_{i\in C}x_i$.

An imputation, called an allocation below, satisfies $x(N)=v(N)$ and $x_i\ge v(\{i\})$ for every $i\in N$. The excess of $C$ at $x$ is $e(C,x)=x(C)-v(C)$. The core is
\[
 \Core(N,v)=\{x\in\mathbb R^N:x(N)=v(N),\ x(C)\ge v(C)\text{ for every }C\subseteq N\}.
\]
Let $X_0$ be the set of allocations and suppose $X_0\ne\varnothing$. For $x\in X_0$, let $\theta(x)$ be the vector obtained by arranging the excesses of the nonempty proper coalitions in nondecreasing order. The \emph{nucleolus}~\cite{Schmeidler} is the unique allocation $\eta\in X_0$ satisfying
\[
 \theta(\eta)\ge_{\mathrm{lex}}\theta(x)\qquad(x\in X_0),
\]
where $\ge_{\mathrm{lex}}$ denotes the lexicographic order.

Maschler's scheme computes the nucleolus by a sequence of linear programs~\cite{MPS}. For a nonempty set $X$ of allocations, define the family of coalitions whose payoff is fixed on $X$ by
\[
 \Fix(X)=\{C\subseteq N:x(C)=y(C)\text{ for all }x,y\in X\}.
\]
Set $\mathcal F_1=2^N$. The empty coalition and the grand coalition both have excess zero at every allocation. At stage $r=1,2,\ldots$, solve
\begin{equation}\label{eq:mps}
 (\mathrm{LP}_r)\qquad
 \begin{alignedat}{2}
  \max_{x,\varepsilon}\quad &\varepsilon\\
  \text{s.t.}\quad &x(C)\ge v(C)+\varepsilon &&\qquad(C\in\mathcal F_r),\\
                   &x\in X_{r-1}.
 \end{alignedat}
\end{equation}
Let $\varepsilon_r$ be the optimal value and $X_r$ the set of allocations $x$ for which $(x,\varepsilon_r)$ is optimal. If $X_r$ is a singleton, its member is the nucleolus. Otherwise set $\mathcal F_{r+1}=2^N\setminus\Fix(X_r)$ and proceed to the next program.

\subsection{The \texorpdfstring{$S$-$T$}{S-T} transversal game}
Let $G=(V,E)$ be a finite undirected graph without loops. We allow parallel edges. Fix a partition $V=S\mathbin{\dot\cup}T$ with both classes nonempty. An edge with one end in each class is a \emph{cross edge}. Put $n=|V|$ and $m=|E|$. A set $C\subseteq E$ is called an \emph{$S$-$T$ connector} if every connected component of the spanning subgraph $(V,C)$ meets both $S$ and $T$. For $U\subseteq V$, let $\delta_G(U)$ be the set of edges with exactly one end in $U$. Consider the cut family
\[
 \mathcal H=\{\delta_G(U):\varnothing\ne U\subseteq S
                         \text{ or }\varnothing\ne U\subseteq T\}.
\]
A transversal of a family of sets meets every member of the family. An edge set $C$ is a connector if and only if $C$ is a transversal of $\mathcal H$. Indeed, a component lying entirely in one class gives a cut missed by $C$. Conversely, if $C$ misses $\delta_G(U)\in\mathcal H$, then $U$ is a nonempty union of components of $(V,C)$, each lying in the same class. Let $\Bcal$ be the family of inclusion minimal connectors, equivalently the minimal transversals of $\mathcal H$. For $A\subseteq E$, write $\chi^A$ for its incidence vector. The \emph{$S$-$T$ transversal game} has player set $E$. The value of a coalition $C\subseteq E$ is
\[
 v(C)=\max\{j:C\text{ contains }j\text{ pairwise disjoint connectors}\}.
\]
Let $k$ denote the maximum number of pairwise disjoint $S$-$T$ connectors in $G$. We assume throughout that $n\ge3$ and $k\ge1$. The other cases have a unique allocation, paying each edge zero if $k=0$ and one if $n=2$. For $n\ge3$, every singleton coalition has value zero, so the allocations are the nonnegative vectors $x$ satisfying $x(E)=k$. We write $\Core$ for $\Core(E,v)$.

\subsection{The packing and transversal linear programs}
Let $\rho$ be the maximum value of a fractional connector packing. Its linear programming dual is the \emph{minimum fractional transversal problem} for the family $\Bcal$ of minimal connectors. A feasible dual vector assigns nonnegative prices to the edges so that each minimal connector has total price at least one. The two programs are
\begin{equation}\label{eq:duality}
\begin{alignedat}{2}
 \max_{\lambda}\quad &\sum_{B\in\Bcal}\lambda_B\\
 \text{s.t.}\quad &\sum_{\substack{B\in\Bcal\\e\in B}}\lambda_B\le1 &&\quad(e\in E),\\
                  &\lambda_B\ge0 &&\quad(B\in\Bcal).
\end{alignedat}
\qquad\qquad
\begin{alignedat}{2}
 \min_y\quad &y(E)\\
 \text{s.t.}\quad &y(B)\ge1 &&\quad(B\in\Bcal),\\
                  &y_e\ge0 &&\quad(e\in E).
\end{alignedat}
\end{equation}

Write $Y$ for the feasible region of the dual and $Y^*=\{y\in Y:y(E)=\rho\}$ for its optimal face. Let $\mathcal D=\conv\{\chi^B:B\in\Bcal\}+\mathbb R^E_+$ be the dominant of the connector polytope. For $i\in\{S,T\}$, form the graph $H_i=G/i$ by identifying all vertices of class $i$ with a single vertex. Loops retain their original labels. For a partition $\Pi$ of the vertices of $H$, let $\delta_H(\Pi)$ denote the edges joining different blocks. The strength of $H$ is
\[
 \sigma(H)=\min_{\Pi:\,|\Pi|\ge2}\frac{|\delta_H(\Pi)|}{|\Pi|-1}.
\]
The capacitated packing theorem of Keijsper and Schrijver~\cite[Theorem~2]{KS} gives $\rho=\min\{\sigma(H_S),\sigma(H_T)\}$.
An edge set $C$ is a connector if and only if $C$ contains a spanning tree in each contracted graph. More generally, for every coalition $C$ and integer $j\ge1$, the packing theorem~\cite{KS} gives
\begin{equation}\label{eq:KSintro}
 v(C)\ge j\quad\Longleftrightarrow\quad
 |C\cap\delta_{H_i}(\Pi)|\ge j(|\Pi|-1)
 \quad(i\in\{S,T\},\ |\Pi|\ge2)
\end{equation}
and hence $k=\lfloor\rho\rfloor$. Applying the same theorem to $(V,C)$ shows that $v(C)$ is the floor of the minimum fractional transversal value for the family $\{B\in\Bcal:B\subseteq C\}$.

A nonempty subpartition $\mathcal W$ of $S$ or $T$ consists of pairwise disjoint nonempty subsets of that class. Put $D(\mathcal W)=\bigcup_{W\in\mathcal W}\delta_G(W)$. For any partition $\Pi$ of $H_i$ with $|\Pi|\ge2$, deleting the block containing the contracted vertex gives a subpartition $\mathcal W$ of the opposite class. Conversely, adding the remaining vertices as a single block recovers $\Pi$. In particular, $\delta_{H_i}(\Pi)=D(\mathcal W)$ and $|\Pi|-1=|\mathcal W|$.

Let $Y_i$ be the nonnegative edge prices giving each spanning tree of $H_i$ total price at least one. The blocking polyhedron of the connectors~\cite[Section~3]{KS} and the partition description for each spanning tree family~\cite{BB} give
\[
 Y=\conv(Y_S\cup Y_T)
  =\conv\left\{\frac{\chi^{D(\mathcal W)}}{|\mathcal W|}:
              \mathcal W\text{ is a nonempty subpartition of }S\text{ or }T\right\}
       +\mathbb R^E_+ .
\]
Minimizing $y(E)$ retains exactly the generators of total price $\rho$. Thus
\begin{equation}\label{eq:dual-optimal-face}
 Y^*=\conv\left\{\frac{\chi^{D(\mathcal W)}}{|\mathcal W|}:
          |D(\mathcal W)|=\rho|\mathcal W|\right\},
\end{equation}
where $\mathcal W$ ranges over the same nonempty subpartitions. Every vertex of $Y^*$ is one of these normalized cut vectors. More precisely, a displayed vector with support $D$ is a vertex if and only if $\chi^D$ lies outside the cone generated by the incidence vectors of the proper subsets of $D$ occurring as supports in~\eqref{eq:dual-optimal-face}. Indeed, any convex representation uses only supports contained in $D$, and normalizing their incidence vectors to the common total price $\rho$ converts a conic representation into a convex one.

\section{The core}\label{sec:core}

\begin{lemma}\label{lem:core-transversal}
For the $S$-$T$ transversal game $(E,v)$, the core is given by
\[
 \Core=\{x\in\mathbb R^E_+: x(E)=v(E),\ x(B)\ge1\text{ for every }B\in\Bcal\}.
\]
\end{lemma}
\begin{proof}
Every minimal connector has value one, so every core allocation satisfies the displayed constraints. Conversely, let $x$ satisfy these constraints and let $C\subseteq E$ have value $j$. If $j=0$, nonnegativity gives $x(C)\ge0$. If $j\ge1$, choose $j$ disjoint minimal connectors $B_1,\ldots,B_j$ in $C$. Then
\[
 x(C)\ge\sum_{i=1}^j x(B_i)\ge j=v(C).
\]
Thus $x$ satisfies every coalition inequality.
\end{proof}

Call a nonempty subpartition $\mathcal W$ of either class \emph{tight} if $|D(\mathcal W)|=k|\mathcal W|$.

\begin{theorem}\label{thm:core-blocker}
The core is nonempty if and only if $k=\rho$. In this case,
\begin{equation}\label{eq:optimal-blocker-hull}
 \Core=\conv\left\{\frac{\chi^{D(\mathcal W)}}{|\mathcal W|}:
       \mathcal W\text{ is a nonempty tight subpartition of }S\text{ or }T\right\}.
\end{equation}
\end{theorem}
\begin{proof}
Lemma~\ref{lem:core-transversal} gives $\Core=\{y\in Y:y(E)=k\}$. The minimum of $y(E)$ over $Y$ is $\rho$, and $k\le\rho$. Hence the core is nonempty precisely when $k=\rho$, in which case $\Core=Y^*$. Formula~\eqref{eq:dual-optimal-face} gives the asserted convex hull.
\end{proof}

\begin{theorem}\label{thm:core-algorithms}
For the $S$-$T$ transversal game, the following problems can be solved in polynomial time.
\begin{enumerate}
 \item Decide whether the core is nonempty.
 \item Find a core allocation when the core is nonempty.
 \item Decide whether a given rational vector belongs to the core.
\end{enumerate}
\end{theorem}
\begin{proof}
Compute a minimum ratio partition in each of $H_S$ and $H_T$~\cite{Gus,Sch}. The smaller of the two ratios is $\rho$. Since $k=\lfloor\rho\rfloor$, Theorem~\ref{thm:core-blocker} shows that the core is nonempty precisely when $\rho$ is an integer. In that case a partition attaining $\rho$ gives a tight subpartition $\mathcal W$, and $\chi^{D(\mathcal W)}/|\mathcal W|$ is a core allocation.

For membership, first check $x\ge0$ and $x(E)=k$. If both hold, find a connector $B$ of minimum weight under $x$. The complements of connectors are the common independent sets of the dual graphic matroids of $H_S$ and $H_T$. Hence a minimum weight connector is the complement of a maximum weight common independent set, which can be found by weighted matroid intersection~\cite{FrankIntersection,KS,Sch}. Since $x\ge0$, some minimum weight connector is minimal under inclusion. By Lemma~\ref{lem:core-transversal}, $x$ belongs to the core if and only if $x(B)\ge1$.
\end{proof}

\section{Edge blocks and their partial orders}\label{sec:blocks}

Fix a maximum packing of $k$ minimal connectors and choose one of its members $B$. At every $x\in\Core$, the members of the packing each receive one unit and the remaining edges receive zero, by disjointness and efficiency.

\subsection{Characterization of the core by exchanges}\label{subsec:one-connector}

A connector is minimal if and only if each component is a tree containing at least one cross edge and all cross edges of that tree share an endpoint. Indeed, deleting an edge of a cycle preserves all components. If a tree contains two disjoint cross edges, deleting an edge on the path between them leaves a cross edge in each of the two components. The cross edges in each tree are therefore pairwise incident and, by bipartiteness, share an endpoint. Conversely, deleting any edge of such a tree leaves a component without a cross edge.

For $i\in\{S,T\}$, let $J_i$ be the bridges of the connected graph $(V(H_i),B)$. Contracting each component left by the deletion of $J_i$ gives a tree with edge set $J_i$. For $e\in E\setminus B$, let $P_i(e)$ consist of the bridges on the path between the images of the ends of $e$, with $P_i(e)=\varnothing$ if the images coincide. For $f\in J_i$,
\[
 f\in P_i(e)\quad\Longleftrightarrow\quad
 (B\setminus\{f\})\cup\{e\}\text{ spans }H_i.
\]
Minimality gives $J_S\cup J_T=B$. Each internal edge is a bridge in the contraction of the opposite class. If a component has a single cross edge, that edge is a bridge in both contractions. If several cross edges share an endpoint in class $i$, they are bridges in $H_i$ and parallel in the other contraction.

\begin{lemma}\label{lem:one-connector}
Let $x\ge0$ satisfy $x(E)=k$. Then $x\in\Core$ if and only if $x(B)=1$ and there exist nonnegative prices $p^i_f$ for $i\in\{S,T\}$ and $f\in B$ such that
\begin{align}
 p^S_f+p^T_f&=x_f &&(f\in B),\notag\\
 p^i_f&=0 &&(i\in\{S,T\},\ f\in B\setminus J_i),\label{eq:one-bridge-split}\\
 x_e&\ge p^S_f+p^T_g &&(e\in E\setminus B,\ f\in P_S(e)\cup\{0\},\ g\in P_T(e)\cup\{0\}).
 \label{eq:one-bridge-path}
\end{align}
The index $0$ has price $p^S_0=p^T_0=0$.
\end{lemma}

\begin{proof}
Fix $i$ and nonnegative edge prices. The set $B$ has minimum price among the spanning sets of $H_i$ if and only if the edges of $B\setminus J_i$ have price zero and each $e\in E\setminus B$ costs at least as much as each bridge on $P_i(e)$. Deleting an edge of $B\setminus J_i$ or replacing a bridge gives necessity. For sufficiency, take a spanning tree contained in $B$. Its price equals the price of $B$. An omitted edge of $B$ closes a cycle of zero price. For an edge $e$ outside $B$, every edge of positive price on the fundamental path belongs to $P_i(e)$. The minimum spanning tree criterion now gives the claim.

For $x\in\Core$, the connector $B$ has price one, the minimum over all connectors by Lemma~\ref{lem:core-transversal}. The complements of connectors are the common independent sets of the duals of the two graphic matroids, so $I=E\setminus B$ maximizes $x$ over these sets. The weight splitting theorem for matroid intersection gives $x=p^S+p^T$ such that $I$ maximizes $p^i$ over the independent sets of the corresponding dual matroid~\cite{FrankIntersection,Sch}. Deleting an element from $I$ shows that $p^i_e\ge0$ for $e\in I$. For $f\notin I$, local optimality requires $p^i_f\le p^i_e$ whenever $I-e+f$ is independent, and $p^i_f\le0$ whenever $I+f$ is independent. Replacing each negative $p^i_f$ by zero preserves these inequalities. For each $f\notin I$ the two resulting prices sum to at least $x_f$ and can be decreased until the sum equals $x_f$, preserving nonnegativity and every inequality. By the exchange criterion, $I$ remains optimal for the resulting nonnegative prices. Applying the preceding spanning set criterion on each side yields \eqref{eq:one-bridge-split} and \eqref{eq:one-bridge-path}.

Conversely, let $x$ and the prices on $B$ satisfy the stated conditions. For each $e\notin B$, assign to each side the largest bridge price on $P_i(e)$, or zero if $P_i(e)$ is empty. Inequality \eqref{eq:one-bridge-path} allows the remaining part of $x_e$ to be distributed between the two sides. Then $B$ is a spanning set of minimum price in each $H_i$. Every connector $C$ spans both sides, so $x(C)\ge p^S(B)+p^T(B)=x(B)=1$. Lemma~\ref{lem:core-transversal} gives $x\in\Core$.
\end{proof}

For each $e\notin B$, introduce two nonnegative variables, one for each path $P_i(e)$. Bound each variable below by the price of every bridge on its path, and bound their sum above by $x_e$. This gives an extended formulation of~\eqref{eq:one-bridge-path} with $O(m^2)$ inequalities. For $f\in J_S\cap J_T$, the prices $p^S_f$ and $p^T_f$ are the contributions of the two spanning requirements to the edge payoff $x_f$.

\begin{example}\label{ex:one-connector-square}
Let $G=K_{2,2}$ with $S=\{s_1,s_2\}$, $T=\{t_1,t_2\}$, and write $a,b,c,d$ for the edges $s_1t_1,s_2t_2,s_1t_2,s_2t_1$, respectively.
The two perfect matchings form a maximum packing. Choose $B=\{a,b\}$. Both edges of $B$ are bridges in both contractions. Set $u_a=p^S_a$ and $u_b=p^S_b$. The bridge paths for $c$ and $d$ give
\[
 \begin{aligned}
 x_c&\ge u_b+x_a-u_a,\\
 x_d&\ge u_a+x_b-u_b .
 \end{aligned}
\]
Their sum gives $x_c+x_d\ge x_a+x_b=1$, the constraint for the other perfect matching $\{c,d\}$. Replacing only $a$ or only $b$ by $c$ leaves an isolated vertex.
\end{example}

\subsection{Closed sets and the computation of a core allocation}\label{subsec:connector-closures}

For $i\in\{S,T\}$ form a directed graph on $J_i\cup(E\setminus B)$. Add an arc from $f$ to $e$ whenever $f\in P_i(e)$. A vertex set $U$ is closed if $f\in U$ implies $e\in U$ for every such arc.

\begin{lemma}\label{lem:connector-closures}
The core is the convex hull of the allocations
\begin{equation}\label{eq:one-closure-allocation}
 \frac{k}{|U|}\chi^U ,
\end{equation}
where $U$ is closed in one of the two directed graphs, $U\cap B\ne\varnothing$, and $|U|=k|U\cap B|$.
\end{lemma}

\begin{proof}
For each closed $U$, its indicator is an admissible price vector on the corresponding side, extended by zero on $B\setminus J_i$. The spanning set criterion in the proof of Lemma~\ref{lem:one-connector} gives $|C\cap U|\ge |B\cap U|$ for every spanning set $C$ of the corresponding $H_i$, hence for every connector. If $B\cap U\ne\varnothing$, division by $|B\cap U|$ gives a feasible vector of \eqref{eq:duality}. In particular,
\[
 |U|\ge k|B\cap U|.
\]
For a nonempty closed set disjoint from $B$ the inequality is strict. If equality holds and $B\cap U\ne\varnothing$, the normalized vector has total price $k$ and equals the allocation in~\eqref{eq:one-closure-allocation}.

Conversely, take $x\in\Core$ and split $x$ into the two prices from the proof of Lemma~\ref{lem:one-connector}. Each side price is nonnegative and nondecreasing along the arcs. Its positive level sets are closed. Taking differences between consecutive distinct price levels expresses $x$ as a sum $\sum_U t_U\chi^U$ with $t_U>0$. The identities $x(E)=k$ and $x(B)=1$ give
\[
 0=\sum_U t_U\bigl(|U|-k|U\cap B|\bigr).
\]
Every summand is nonnegative, so all the level sets in this sum satisfy equality. The coefficients $t_U|U\cap B|$ sum to one. They give the claimed convex combination.
\end{proof}

\paragraph{Maximum weight closed sets.}
Give each node $e$ of the exchange graph weight $k\chi^B_e-1$. A closed set $U$ then has weight $k|U\cap B|-|U|$, whose maximum is zero.

The usual closure network has an arc from the source to every node of positive weight and an arc from every node of negative weight to the sink, with capacity equal to the absolute weight. Give each exchange arc capacity greater than the sum of the positive weights. The source sides of minimum cuts, with the source removed, are precisely the closed sets of maximum weight~\cite{Picard,PQ}. Thus any nonempty such set on either side gives the allocation in \eqref{eq:one-closure-allocation}.

\paragraph{The order on edge blocks.}

For $B$ as in Lemma~\ref{lem:one-connector} and $i\in\{S,T\}$, let $\mathcal U_i$ consist of the closed sets on side $i$ satisfying $|U|=k|U\cap B|$, including the empty set. An edge is active on side $i$ if some member of $\mathcal U_i$ contains the edge. Two active edges belong to the same block if they belong to exactly the same members of $\mathcal U_i$. For active blocks $Q,R$ on side $i$, write $Q\preceq_i R$ if every member of $\mathcal U_i$ containing $Q$ also contains $R$. Let $0$ denote the block of inactive edges, which may be empty. Write $[e]_i$ for the block of $e$ on side $i$. Write $Q\to R$ for a cover of the order.

\begin{theorem}\label{thm:residual-blocks}
The members of $\mathcal U_i$ are exactly the unions of active blocks that are upper sets of $\preceq_i$. A vector $x\in\mathbb R^E$ belongs to the core if and only if there exist block prices satisfying
\[
 \begin{aligned}
 x(E)&=k, & x_e&=p^S_{[e]_S}+p^T_{[e]_T} &&(e\in E),\\
 p^i_0&=0, & p^i_Q&\ge 0 &&(Q\text{ active on side }i),\\
 &&p^i_Q&\le p^i_R &&(Q\to R\text{ a cover on side }i).
 \end{aligned}
\]
Given the maximum packing, one maximum flow in each closure network suffices to find the blocks and covers.
\end{theorem}

\begin{proof}
The closed sets on each side are closed under union and intersection. The function $U\mapsto k|U\cap B|-|U|$ is modular and has maximum zero. Its maximizers therefore also form a family closed under union and intersection. For an active edge $e$, the intersection of all maximizers containing $e$ is the union of the blocks above $[e]_i$, including $[e]_i$. Unions of these principal upper sets give all upper sets and belong to $\mathcal U_i$. The defining implication of the order gives the converse.

Nonnegative prices that respect the cover inequalities are nonnegative combinations of upper set indicators, as seen by taking the sets of blocks at or above each positive price. Conversely, every such indicator respects the cover inequalities. The sum of the two order cones is therefore generated by the vectors $\chi^U$ for nonempty $U\in\mathcal U_S\cup\mathcal U_T$. Intersecting this cone with $x(E)=k$ gives the convex hull in Lemma~\ref{lem:connector-closures}.

The empty closed set is optimal, so the cut with only the source on its source side is minimum. After a maximum flow, a minimum cut has a source side closed under residual arcs. Nodes that can reach the sink belong to no optimal closed set. Among the remaining edge nodes, reachability gives the implications of $\preceq_i$. Indeed, if $f$ cannot reach $e$, the residual nodes reachable from $f$, together with the source, form a minimum cut source side containing $f$ and excluding $e$. Thus the strongly connected components on the active edge nodes give the blocks. Edges omitted from the exchange graph of side $i$ are inactive on that side.
\end{proof}

\paragraph{Testing core nonemptiness.}
By Lemma~\ref{lem:connector-closures}, the core is nonempty if and only if one of the two closure networks has a nonempty closed set of maximum weight, or equivalently, one side has an active block. In this case, the residual graph gives a nonempty tight closed set $U$, and~\eqref{eq:one-closure-allocation} gives a core allocation.

\subsection{The block incidence graph}\label{subsec:block-incidence}

For each side $i$, let $\mathscr B_i$ consist of its active blocks and the ground block $0$. Form a directed graph on $\mathscr B_i$ with an arc for each cover of $\preceq_i$ and an arc from $0$ to every active block. Write $\mathscr A_i$ for its arc set. The admissible side prices in Theorem~\ref{thm:residual-blocks} form the cone
\[
 K_i=\{p^i:p^i_0=0,\ p^i_C\ge p^i_B\text{ for every }BC\in\mathscr A_i\}.
\]
For a vector $p^i$ of active block prices, set $p^i_0=0$ and define the linear map $L_i$ by $(L_ip^i)_e=p^i_{[e]_i}$. Thus the two block prices contributing to $x_e$ are attached to the endpoints of $e$ in the following graph.

Form an undirected multigraph $J$ on the active blocks of both sides and a common ground vertex $0$. Each original edge $e$ joins $[e]_S$ to $[e]_T$, identifying the two ground blocks. Call the component of $0$ the \emph{root component} and the remaining components \emph{unrooted}. Every unrooted component contains active blocks from both sides.

In $H_i$, each cross edge is incident with the contracted vertex, called the root. Each internal edge is either a root loop or has both endpoints away from the root. An active block containing only cross edges is a \emph{pure crossing block}.

\begin{lemma}\label{lem:pure-cross-blocks}
Every block in an unrooted component of $J$ is a pure crossing block. On either side, no arc joins two distinct pure crossing blocks. In particular, every arc of $\mathscr A_i$ has at least one end in the root component of $J$.
\end{lemma}

\begin{proof}
Fix a side with an active block and take disjoint spanning trees $T_1,\ldots,T_k$ in $H_i$, one from each connector of the fixed packing. On $w(E)=k$, the cone $L_iK_i$ consists of the nonnegative prices satisfying $w(T)\ge1$ for every spanning tree $T$ of $H_i$. Indeed, for each tight closed set $U$, the spanning set criterion in Lemma~\ref{lem:one-connector} gives $|T\cap U|\ge|B\cap U|=|U|/k$. Conversely, if every tree costs at least one, the disjoint connectors of the packing each cost one. Hence $B$ is a minimum price spanning set in $H_i$. Its bridge inequalities make every positive level set of $w$ closed, and the equality $w(E)=kw(B)$ makes each such set tight by Lemma~\ref{lem:connector-closures}.

An active block belongs to a nonempty tight closed set, whose indicator can be scaled to give $w(E)=k$. Each selected tree costs at least one and $w(E)=k$, so each costs one, all unused edges cost zero, and the $k$ trees form a maximum packing. By the graphic case of the exchange characterization~\cite[Theorem~2 and Corollary~1]{MP}, the active blocks are the strongly connected components of the tree exchange graph on edges of potentially positive price. Their order is given by reachability, so each cover between active blocks is induced by an exchange.

Give the active blocks distinct positive prices increasing along a linear extension, and scale them so that $w(E)=k$. Let $e$ belong to a pure crossing block $Q$ of price $\alpha$, and let $T_j$ omit $e$. The path $P$ in $T_j$ joining the ends of $e$ starts at the root. All its edges cost at most $\alpha$ by optimality of $T_j$. Some edge of $P$ costs $\alpha$, since otherwise a selected tree containing $e$ could replace $e$ by a cheaper edge of $P$. Every edge of price $\alpha$ belongs to $Q$ and meets the root, so the first edge of $P$ is its only edge of price $\alpha$. All later edges of $P$ are internal. Thus any exchange into $Q$ from a different active block has an internal edge at its tail.

An internal edge of an active block is a loop in the other contraction. Its incidence edge in $J$ reaches the ground vertex. Every block in an unrooted component therefore contains only cross edges. An arc whose ends both lie outside the root component would join two distinct pure crossing blocks on the same side.
\end{proof}

\section{The nucleolus}\label{sec:nonempty-nucleolus}
Throughout this section we assume that the core is nonempty. In the first program of~\eqref{eq:mps}, efficiency gives $e(E,x)=0$ for every allocation, and every core allocation has nonnegative excess for every coalition. Hence $\varepsilon_1=0$ and $X_1=\Core$.

\subsection{Coalitions in the second linear program}\label{subsec:second-coalitions}

A coalition $A$ has constant payoff on the core if and only if the ratios $|A\cap Q|/|Q|$ agree for all active blocks on both sides. Indeed, on each side a linear extension gives strictly increasing positive block prices, so each order cone is full dimensional in its active block coordinates. After normalization, the affine hull in price coordinates is the hyperplane $\sum_{i,Q}|Q|p^i_Q=k$. The payoff to $A$ is $\sum_{i,Q}|A\cap Q|p^i_Q$. Constancy on the hyperplane gives the ratio condition.

By Theorem~\ref{thm:residual-blocks}, the core lies on a single ray precisely when the incidence vectors of the active blocks have rank one. In this case efficiency determines the unique core allocation, which is the nucleolus and can be found by Theorem~\ref{thm:core-algorithms}.

For the rest of this section, suppose the core contains more than one allocation.

\begin{proposition}\label{prop:second-core-coalitions}
The second program in~\eqref{eq:mps} is equivalent to
\[
\begin{alignedat}{2}
 \max_{x,\varepsilon}\quad &\varepsilon\\
 \text{s.t.}\quad &x\in\Core,\\
 &x_e\ge\varepsilon &&\qquad(\{e\}\notin\Fix(\Core)),\\
 &x(C)\ge1+\varepsilon &&\qquad(C\in\Bcal,\ C\notin\Fix(\Core)).
\end{alignedat}
\]
\end{proposition}
\begin{proof}
Take a maximum packing of minimal connectors in a coalition $C$. The excess of $C$ is the sum of the connector excesses and the payoffs to the remaining edges. Each summand is nonnegative on the core. If $C$ is unfixed, at least one summand varies, and its singleton or connector inequality gives $e(C,x)\ge\varepsilon$. Conversely, each displayed inequality is a coalition constraint of $(\mathrm{LP}_2)$.
\end{proof}

Let $\one$ denote the vector of ones in $\mathbb R^E$. The sum of the incidence vectors of a maximum packing is at most $\one$, so $\one\in k\mathcal D$. To express the coalition excesses in block prices, consider the integer directions from $\one$ in $k\mathcal D$. By Theorem~\ref{thm:residual-blocks}, the cone generated by the core is $L_SK_S+L_TK_T$.

Let $\mathcal T$ be the tangent cone of $k\mathcal D$ at $\one$, defined by
\[
 \mathcal T=\{z:\ \one+t z\in k\mathcal D
                    \text{ for all sufficiently small }t>0\}.
\]
On the active blocks, the transpose of $L_i$ satisfies
\[
 (L_i^\top z)_B=\sum_{e\in B}z_e\qquad(B\in\mathscr B_i\setminus\{0\}).
\]
For an arc $a=BC\in\mathscr A_i$ of the block digraph, let $d_a$ be its incidence column, with $1$ at the head $C$, $-1$ at the tail $B$, and the coordinate of the ground block omitted. Then
\begin{equation}\label{eq:integer-tangent-flow}
 z\in\mathcal T\quad\Longleftrightarrow\quad
 L_i^\top z\in\operatorname{cone}\{d_a:a\in\mathscr A_i\}
             \quad(i\in\{S,T\}).
\end{equation}
The cone of minimizing objectives of $k\mathcal D$ at $\one$ is $\operatorname{cone}(\Core)$. Indeed, a nonzero minimizing objective $w$ is nonnegative, and $k w(B)\ge w(E)$ for each $B\in\Bcal$ since $k\chi^B\in k\mathcal D$. Thus $kw/w(E)\in\Core$. Conversely, for $x\in\Core$ every point of $k\mathcal D$ costs at least $k=x(E)$ under $x$. Thus $z\in\mathcal T$ precisely when $(L_i^\top z)^\top p^i\ge0$ for every $p^i\in K_i$ on each side. Farkas' lemma gives \eqref{eq:integer-tangent-flow}. For integral $z$, the flows on the right can be chosen integral by the integral transshipment theorem~\cite[Corollary~11.2f]{Sch}.

\begin{lemma}\label{lem:tangent-coalition-slacks}
The following statements hold on the entire core.
\begin{enumerate}
 \item For every coalition $A\subseteq E$, there exists $z_A\in\mathcal T\cap\mathbb Z^E$ such that
 \[
             e(A,x)=x^\top z_A\qquad(x\in\Core).
 \]
 \item For every $z\in\mathcal T\cap\mathbb Z^E$, the function $x^\top z$ is a finite nonnegative integer linear combination of coalition excesses.
\end{enumerate}
\end{lemma}

\begin{proof}
Fix a maximum packing $B_1,\ldots,B_k\in\Bcal$ in the full graph. At each $x\in\Core$, the packing members have price one and every edge outside their union has price zero.
For a coalition with $v(A)=j$, set $z_A=\chi^A-\sum_{\ell=1}^j\chi^{B_\ell}$, where the sum is empty if $j=0$. Replace the first $j$ connectors of the fixed packing by $j$ connectors contained in $A$. Their incidence vectors, together with those of the remaining $k-j$ connectors, are dominated by $\one+z_A$. Hence $\one+z_A\in k\mathcal D$, so $z_A\in\mathcal T$ and $x^\top z_A=x(A)-j$.

Conversely, given an integer vector $z\in\mathcal T$, choose a sufficiently large positive integer $q$ such that
\[
                  q\one+z\in qk\mathcal D\cap\mathbb Z_+^E.
\]
The capacitated packing theorem of Keijsper and Schrijver~\cite[Theorem~2]{KS} gives a packing of $qk$ connectors under the capacities $q\one+z$. Taking a minimal connector in each gives $F_1,\ldots,F_{qk}\in\Bcal$ and an integral vector of unused capacities $c\ge0$ satisfying
\[
               q\one+z=\sum_{\ell=1}^{qk}\chi^{F_\ell}+c.
\]
Minimality implies $v(F_\ell)=1$, and $n\ge3$ ensures that $v(\{e\})=0$. Using $x(E)=k$, we obtain
\begin{equation}\label{eq:tangent-slack-decomposition}
 x^\top z=\sum_{\ell=1}^{qk}e(F_\ell,x)
                   +\sum_{e\in E}c_e e(\{e\},x).
\end{equation}

\end{proof}

\subsection{Directed cycles and coalition excesses}

To combine the two flow conditions in~\eqref{eq:integer-tangent-flow}, add the arcs of $\mathscr A_S$ and the reversals of the arcs of $\mathscr A_T$ to $J$. Shrink each component of $J$ to one vertex and retain the directed arcs. Denote the resulting directed multigraph by $\Omega$ and the vertex corresponding to the root component by $0$. Give the image of each original arc $a=BC\in\mathscr A_i$ cost
\[
                  \Delta_a(p)=p^i_C-p^i_B.
\]
The cost uses the original direction on each side. By Lemma~\ref{lem:pure-cross-blocks}, every arc of $\Omega$ is a loop at $0$ or joins $0$ to an unrooted component. Hence the simple directed cycles are
\begin{enumerate}
 \item a loop at the ground vertex, corresponding to one arc of the block digraph and having cost $\Delta_a$,
 \item for an unrooted component $U$, an arbitrary arc $a$ entering $U$ together with an arbitrary arc $b$ leaving $U$, having total cost $\Delta_a+\Delta_b$.
\end{enumerate}
Call these \emph{elementary cycles}, denote their family by $\mathscr R$, and write $r_R(p)$ for the cost of $R$. Each unrooted component has an entering arc and a leaving arc, since each side has a ground arc to every active block. The number of elementary cycles is at most the square of the number of block arcs.

\begin{lemma}\label{lem:ring-integer-decomposition}
For every elementary cycle $R$, there exists an integer direction $z_R\in\mathcal T$ such that
\[
             r_R(p)=x^\top z_R
             \quad\text{whenever }x=\sum_{i\in\{S,T\}}L_ip^i.
\]
Conversely, for each integral $z\in\mathcal T$, the function $x^\top z$ on the core is a nonnegative integer linear combination of the costs of elementary cycles.
\end{lemma}

\begin{proof}
Orient the incidence edges of $J$ from $S$ to $T$ and give each edge $e$ signed flow $z_e$. At the active blocks the net inflows are $-L_S^\top z$ and $L_T^\top z$. For integral $z\in\mathcal T$, the nonnegative integral flows in \eqref{eq:integer-tangent-flow} cancel these net inflows once the $T$ arcs are reversed. Conservation at $0$ follows by summing over all other vertices. Contracting the incidence components gives a circulation on $\Omega$.

Conversely, a nonnegative integral circulation on $\Omega$ gives integer imbalances summing to zero in each incidence component before contraction. Route these imbalances along a spanning tree of each component of $J$, using signed integral flows $z_e$. The resulting $z$ satisfies \eqref{eq:integer-tangent-flow}. For a unit circulation around $R$, denote this vector by $z_R$. Summing block prices against the net inflows in the original block digraphs gives $x^\top z_R=r_R(p)$.

For any integral $z\in\mathcal T$, decompose the induced circulation on $\Omega$ into unit circulations around simple directed cycles~\cite[Theorem~11.1]{Sch}. Subtracting their vectors $z_R$ leaves a vector with sum zero on every active block. Its inner product with every core allocation is zero, which proves the converse.
\end{proof}

By Lemma~\ref{lem:ring-integer-decomposition}, the cost of each elementary cycle $R$ depends only on the allocation. Write $r_R(x)=x^\top z_R$ for this cost.

\begin{proposition}\label{prop:ring-exact-minimum}
Each elementary cycle has nonconstant cost on the core. For every $x\in\Core$,
\begin{equation}\label{eq:ring-min-equality}
 \min_{A\notin\Fix(\Core)}e(A,x)=\min_{R\in\mathscr R}r_R(x).
\end{equation}
\end{proposition}

\begin{proof}

Take a point in the relative interior of each $K_i$ and scale jointly to total price $k$. Every arc has strictly positive cost there, so any constant cycle cost is positive.

Suppose $r_R$ is constant. By the affine hull of the normalized block prices in Section~\ref{subsec:second-coalitions}, the coefficient of $p^i_B$ in $r_R$ is $\lambda|B|$ for a common $\lambda>0$ and every active $B$.

With active blocks on just one side, $R$ consists of a single arc. Only its head has positive coefficient. Thus the side has precisely one active block, and $x(E)=k$ fixes its price.

With active blocks on both sides, all their coefficients must be positive. Thus $R$ uses one arc from each side, each starting at its original ground block, and its head is the only active block on that side. These two active blocks lie in the same unrooted component. Since every incidence edge at one ends at the other, they coincide as subsets of $E$. Their lifts span one ray, on which $x(E)=k$ again fixes the allocation.

Both cases give a singleton core, a contradiction.

By Lemmas~\ref{lem:tangent-coalition-slacks} and~\ref{lem:ring-integer-decomposition}, each excess is a nonnegative integer sum of cycle costs. A nonconstant excess has a nonconstant summand, and every other summand is nonnegative on the core. This gives the lower bound in \eqref{eq:ring-min-equality}.

For the reverse inequality, each cycle cost is a nonnegative integer sum of coalition excesses by the same lemmas. A nonconstant cost has a nonconstant summand, and all excesses are nonnegative on the core. This gives the reverse bound. Both minima are over nonempty families, since some singleton payoff varies on the core.
\end{proof}

\subsection{Block prices and the nucleolus}

The two types of elementary cycle in $\Omega$ determine the arc lengths used to compute the nucleolus. For each arc $a\in\mathscr A_i$ of the block digraph, define its length by
\begin{equation}\label{eq:arc-length-half}
 \ell_a=\begin{cases}
  1,&\text{if its image in }\Omega\text{ is a loop at $0$},\\
  \tfrac12,&\text{if its image joins $0$ to an unrooted component}.
 \end{cases}
\end{equation}

Suppose two pairs of block prices give the same allocation $x$. Their difference vanishes on the root component. On each unrooted component $U$ the pairs differ by a shift
\begin{equation}\label{eq:unanchored-gauge}
       (p^S_B,p^T_C)\longmapsto(p^S_B+c_U,p^T_C-c_U)
       \qquad(B\in U\text{ on side }S,\ C\in U\text{ on side }T).
\end{equation}
The price differences at the ends of each incidence edge sum to zero. They therefore vanish on the root component and alternate in sign along paths in each unrooted component. Each such shift preserves $x$.

\begin{lemma}\label{lem:second-prices}
Let $x\in\Core$ and $\varepsilon\ge0$. The allocation $x$ satisfies the coalition constraints of Proposition~\ref{prop:second-core-coalitions} at $\varepsilon$ if and only if $x=L_Sp^S+L_Tp^T$ for some $p^i\in K_i$ satisfying
\begin{equation}\label{eq:lifted-length-constraints}
 p^i_C-p^i_B\ge\varepsilon\ell_a\qquad(a=BC\in\mathscr A_i,\ i\in\{S,T\}).
\end{equation}
\end{lemma}

\begin{proof}
By Proposition~\ref{prop:ring-exact-minimum}, the coalition constraints hold precisely when every elementary cycle costs at least $\varepsilon$. If the prices satisfy~\eqref{eq:lifted-length-constraints}, summing around each elementary cycle gives this bound, since its arc lengths sum to one.

Conversely, suppose all elementary cycles cost at least $\varepsilon$ and fix block prices representing $x$. A loop already has cost at least $\varepsilon$. At each unrooted component $U$, let $\alpha_U$ be the minimum entering cost and $\beta_U$ the minimum leaving cost. The cycle inequalities at $U$ give
\[
                         \alpha_U+\beta_U\ge\varepsilon.
\]
Under the shift \eqref{eq:unanchored-gauge}, entering costs increase by $c_U$ and leaving costs decrease by $c_U$, because the $T$ arcs in $\Omega$ are reversed. Choose
\[
               \frac{\varepsilon}{2}-\alpha_U
               \ \le c_U\le
               \beta_U-\frac{\varepsilon}{2}.
\]
Such a choice exists by the preceding inequality. The choices are independent across unrooted components and preserve $x$. Every incident arc then has cost at least $\varepsilon/2$. All shifted prices satisfy \eqref{eq:lifted-length-constraints} and belong to the cones $K_i$.
\end{proof}

For each active block $Q$ on side $i$, let $h^i_Q$ be the length of a longest directed path from $0$ to $Q$ in the block digraph with arc lengths~\eqref{eq:arc-length-half}. Set $h^i_0=0$.

\begin{theorem}\label{thm:body-nonempty}
Put $h_e=h^S_{[e]_S}+h^T_{[e]_T}$. Then $(\mathrm{LP}_2)$ has a unique optimal solution $(\eta,\varepsilon_2)$ given by
\begin{equation}\label{eq:nonempty-nucleolus-formula}
 \begin{aligned}
 \varepsilon_2&=\frac{k}{\sum_{e\in E}h_e},\\
 \eta_e&=\varepsilon_2 h_e\quad(e\in E).
 \end{aligned}
\end{equation}
The algorithm uses one maximum connector packing, two maximum flow computations, and longest paths in two acyclic digraphs, together with graph searches and linear algebra.
\end{theorem}
\begin{proof}
By Lemma~\ref{lem:second-prices}, summing the arc inequalities along a longest path gives $p^i_Q\ge\varepsilon h^i_Q$ in a feasible price representation. Equality for every block satisfies the arc inequalities by the definition of a longest path. Summing over all edges gives $\varepsilon\sum_e h_e\le k$, with equality only if every active block attains its lower bound. Hence~\eqref{eq:nonempty-nucleolus-formula} gives the unique optimal allocation of $(\mathrm{LP}_2)$ and therefore the nucleolus.
\end{proof}

\begin{example}[The triangular prism]\label{ex:prism}
Let $S$ and $T$ be the two triangular faces of a triangular prism, with three edges joining corresponding vertices. Discarding loops gives $K_4$ in either contraction. In $K_4$, a minimum cut has three edges, a partition into three blocks has at least five, and the singleton partition has six. Its strength is therefore $6/(4-1)=2$, attained only by the singleton partition. Thus $k=\rho=2$.

By the partition description in Theorem~\ref{thm:core-blocker}, the tight partition on each side yields a core allocation assigning $1/3$ to the six nonloop edges of that contracted graph and zero to the remaining edges. Thus every core allocation gives a common price $p$ to the edges internal to $S$, a common price $q$ to those internal to $T$, and $p+q$ to each of the three crossing edges, where $p,q\ge0$ and $p+q=1/3$. For a coalition $C$ of value $j$, let $a$ and $b$ be its numbers of nonloop edges in the two contracted graphs, ordered to correspond to the prices $p$ and $q$. Since $a,b\ge3j$,
\[
 e(C,x)=(a-3j)p+(b-3j)q .
\]
If $C$ is unfixed, at least one coefficient is a positive integer, so $e(C,x)\ge\min\{p,q\}$. Internal singleton coalitions attain $p$ and $q$. Thus $(\mathrm{LP}_2)$ maximizes $\min\{p,q\}$ and has the unique solution $p=q=1/6$. The nucleolus assigns $1/6$ to each internal edge and $1/3$ to each cross edge.
\end{example}

\end{document}